\documentclass[12pt,a4paper]{article}
\makeatletter
\def\eqnarray{%
\stepcounter{equation}%
\let\@currentlabel=\theequation
\global\@eqnswtrue
\global\@eqcnt\z@
\tabskip\@centering
\let\\=\@eqncr
$$\halign to \displaywidth\bgroup\@eqnsel\hskip\@centering
$\displaystyle\tabskip\z@{##}$&\global\@eqcnt\@ne
\hfil$\displaystyle{{}##{}}$\hfil
&\global\@eqcnt\tw@$\displaystyle\tabskip\z@{##}$\hfil
\tabskip\@centering&\llap{##}\tabskip\z@\cr}
\makeatother
\newcommand{\kansu}[2]{{{#1}\!\left({#2}\right)}}
\newcommand{\ket}[1]{{\vert{#1}\rangle}}
\newcommand{\bra}[1]{{\langle{#1}\vert}}
\newcommand{\calh}{{\cal H}}
\newcommand{\cala}{{\cal A}}
\newcommand{\calf}{{\cal F}}
\newcommand{\calp}{{\cal P}}
\newcommand{\fukuso}{{\mathbf C}}
\newcommand{\futon}{{\bf N}}

\newcommand{\stm}{{St_m}}
\newcommand{\grm}{{Gr_m}}
\newcommand{\muzeta}{{\vert\mu\vert}}

\begin{document}

\title{\sl Note on Coherent States and Adiabatic Connections, Curvatures}
\author{
  Kazuyuki FUJII
  \thanks{E-mail address : fujii@math.yokohama-cu.ac.jp }\\
  Department of Mathematical Sciences\\
  Yokohama City University\\
  Yokohama, 236-0027\\
  Japan
  }
\date{}
\maketitle\thispagestyle{empty}
%
%
%
%
\begin{abstract}
  We give a possible generalization to the example in the paper of 
  Zanardi and Rasetti (quant--ph 9904011). For this generalized 
  one explicit forms of adiabatic connection, curvature and etc.
  are given.
\end{abstract}

\newpage

%
%
%
%

This is a comment paper to Zanardi and Rasetti\cite{ZM1} and 
the aim is to give a mathematical inforcement to \cite{ZM1}.

After the breakthrough by P. Shor\cite{PS} there has been remarkable
progress in Quantum Computation (or Computer)(QC briefly).
See \cite{LPS} in outline.

On the other hand, Gauge Theories are widely recognized as the basis in 
quantum field theories.
Therefore it is very natural to intend to include  gauge theories 
in QC $\cdots$ a construction of ``gauge theoretic'' quantum computation
or of ``geometric'' quantum computation in our terminology.

Zanardi and Rasetti proposed in \cite{ZM1} and  \cite{ZM2} such an idea 
using non-abelian Berry phase (quantum holonomy), see also \cite{JP}.
In their model a Hamiltonian (including some parameters) must be
degenerated because an adiabatic connection is introduced using
this degeneracy \cite{SW}.

They gave a simple example to explain their idea.
However there are many misprints in their calculations,
so it is not easy to follow their idea.

We believe that this example will become important in
the near future.
Therefore we deal with it once more and give a possible generalization.
For the generalized model explicit forms of adiabatic (Berry) connection,
curvature and etc are given.

It is not easy to predict the future of gauge theoretic quantum computation.
However it is an arena worth challenging for mathematical physicists.

\vspace{5mm}

We start with mathematical preliminaries.
Let $\calh$ be a separable Hilbert space over $\fukuso$.
For $m\in{\bf N}$, we set
\begin{equation}
  \label{eq:stmh}
  \kansu{\stm}{\cal H}
  \equiv
  \left\{
    V=\left( v_1,\cdots,v_m\right)
    \in
    \calh\times\cdots\times\calh\vert V^\dagger V=1_m\right\}\ ,
\end{equation}
where $1_m$ is a unit matrix in $\kansu{M}{m,\fukuso}$.
This is called a (universal) Stiefel manifold.
Note that the unitary group $U(m)$ acts on $\kansu{\stm}{\calh}$
from the right:
\begin{equation}
  \label{eq:stmsha}
  \kansu{\stm}{\calh}\times\kansu{U}{m}
  \rightarrow
  \kansu{\stm}{\calh}: \left( V,a\right)\mapsto Va.
\end{equation}
Next we define a (universal) Grassmann manifold
\begin{equation}
  \kansu{\grm}{\calh}
  \equiv
  \left\{
    X\in\kansu{M}{\calh}\vert
    X^2=X, X^\dagger=X\  \mathrm{and}\  \mathrm{tr}X=m\right\}\ ,
\end{equation}
where $M(\calh)$ denotes a space of all bounded linear operators on $\calh$.
Then we have a projection
\begin{equation}
  \label{eq:piteigi}
  \pi : \kansu{\stm}{\calh}\rightarrow\kansu{\grm}{\calh}\ ,
  \quad \kansu{\pi}{V}\equiv VV^\dagger\ ,
\end{equation}
compatible with the action (\ref{eq:stmsha}) 
($\kansu{\pi}{Va}=Va(Va)^\dagger=Vaa^\dagger V^\dagger=VV^\dagger=
\kansu{\pi}{V}$).

Now the set
\begin{equation}
  \left\{
    \kansu{U}{m}, \kansu{\stm}{\calh}, \pi, \kansu{\grm}{\calh}
  \right\}\ ,
\end{equation}
is called a (universal) principal $U(m)$ bundle, 
see \cite{MN} and \cite{KF}.

Next let $M$ be a $n$-dimensional differentiable manifold and 
the map $P : M \rightarrow \kansu{\grm}{\calh}$ be given.
For this $P$ the pull-back bundle over $M$ is defined as follows\cite{MN}:
\begin{eqnarray}
  \label{eq:hikimodoshi}
  &&\left(\kansu{U}{m}, E, \pi_E, M\right)
  \equiv
  P^*\left(\kansu{U}{m}, \kansu{\stm}{\calh}, \pi, \kansu{\grm}{\calh}\right)
  \ ,
  \nonumber\\
  &&E
  =
  \left\{
    \left( x, V\right)\in M\times\kansu{\stm}{\calh}\vert
    \kansu{P}{x}=\kansu{\pi}{V}
  \right\}\ ,
  \nonumber\\
  &&\pi_E : E\rightarrow M\ ,\ \kansu{\pi_E}{\left( x,V\right)}=x\ .
  \nonumber\\
  \nonumber\\
  &&\matrix{
    \kansu{U}{m}&&\kansu{U}{m}\cr
    \Big\downarrow&&\Big\downarrow\cr
    E&\longrightarrow&\kansu{\stm}{\calh}\cr
    \Big\downarrow&&\Big\downarrow\cr
    M&\longrightarrow&\kansu{\grm}{\calh}\cr
    }
\end{eqnarray}
For the (canonical) local section induced from that of (\ref{eq:piteigi})
\begin{equation}
  \phi : U ( \mathrm{open} \subset M) \rightarrow E\ ,
\end{equation}
we can write $\phi(x)=(x,V(x))$ on $U$, so the canonical $1$-form $\cala$
(gauge field) is defined as
\begin{equation}
  \label{eq:ateigi}
  \cala\equiv V^\dagger dV\ \mathrm{on}\ U\ ,
\end{equation}
where $d$ is a differential form on $U\subset M$.
This is a local form.
From this we obtain a curvature form
\begin{equation}
  \label{eq:fteigi}
  \calf
  \equiv
  d\cala+\cala\wedge\cala
  =
  dV^\dagger\wedge dV+V^\dagger dV\wedge V^\dagger dV\ .
\end{equation}
Now if we define a map $P(x)=V(x)V(x)^\dagger$,
the curvature $2$-form of this (induced) bundle is given by
$PdP\wedge dP$ which is related to (\ref{eq:fteigi}) by
\begin{equation}
  \label{eq:pvr}
  PdP\wedge dP
  =
  V\left( d\cala+\cala\wedge\cala\right) V^\dagger\ .
\end{equation}
The left hand side of (\ref{eq:pvr}) is a global form.

We are very interested in the example in \cite{ZM1},
so we give a possible generalization of that and study it in detail.

Let $a(a^\dagger)$ be the annihilation (creation) operator of the harmonic 
oscillator.
If we set $N\equiv a^\dagger a$ (:\ number operator), then
\begin{equation}
  [N,a^\dagger]=a^\dagger\ ,\
  [N,a]=a\ ,\
  [a,a^\dagger]=1\ .
\end{equation}
Let $\calh$ be a Fock space generated by $a$ and $a^\dagger$, and
$\{\ket{n}\vert n\in\futon\cup\{0\}\}$ be its basis.
The actions of $a$ and $a^\dagger$ on $\calh$ are given by
\begin{equation}
  \label{eq:shoukou}
  a\ket{n} = \sqrt{n}\ket{n-1}\ ,\
  a^\dagger\ket{n} = \sqrt{n+1}\ket{n+1}\ ,
\end{equation}
where $\ket{0}$ is a vacuum ($a\ket{0}=0$).

For $\lambda\in\fukuso$ the coherent state $\ket{\lambda}\in\calh$
is given by
\begin{equation}
  \label{eq:lambdateigi}
  \ket{\lambda} = e^{\lambda a^\dagger-\bar\lambda a}\ket{0} \ ,
\end{equation}
see \cite{KS}.
By the elementary Baker-Campbell-Hausdorff formula\cite{KS},
the unitary operator in (\ref{eq:lambdateigi}) is decomposed into
\begin{equation}
  \label{eq:ktachi}
  e^{\lambda a^\dagger-\bar\lambda a}
  =
  e^{-\vert\lambda\vert^2/2}e^{\lambda a^\dagger}e^{-\bar\lambda a}\ .
\end{equation}
Next we assign
\begin{equation}
  \label{eq:kdaisu}
  K_+\equiv{1\over2}\left( a^\dagger\right)^2\ ,\
  K_-\equiv{1\over2}a^2\ ,\
  K_3\equiv{1\over2}\left( a^\dagger a+{1\over2}\right)^2\ .
\end{equation}
Then we have
\begin{equation}
  [K_3,K_+]=K_+\ ,\
  [K_3,K_-]=-K_-\ ,\
  [K_+,K_-]=-2K_3\ .
\end{equation}
That is, the set $\{K_+,K_-,K_3\}$ gives a unitary representation of $su(1,1)$
with spin $1/4$, \cite{AP}.
For $\mu\in\fukuso$ the squeezed state (the coherent state of Perelomov's
type in our terminology \cite{FS}) $\ket{\mu}$ is given by
\begin{equation}
  \label{eq:muteigi}
  \ket{\mu}
  \equiv
  e^{\mu K_+-\bar\mu K_-}\ket{{1\over4},0}\ ,
\end{equation}
where $\ket{{1\over4},0}$ is a vacuum ($K_-\ket{{1\over4},0}=0$).
Now applying the disentangling formula \cite{AP} and \cite{FS} to the
unitary operator in (\ref{eq:muteigi}) we obtain
\begin{equation}
  \label{eq:bunkai}
  e^{\mu K_+-\bar\mu K_-}
  =
  e^{\zeta K_+}e^{\kansu{\log}{1-\vert\zeta\vert^2}K_3}
  e^{-\bar\zeta K_-}\ ,
\end{equation}
where $\zeta=\mu\tanh\muzeta/\muzeta$.

Under preliminaries above let us proceed to the main subject.
Let $H_0$ be a Hamiltonian
\begin{equation}
  \label{eq:ham}
  H_0\equiv\hbar\omega N(N-1)\cdots(N-m+1)\ ,
\end{equation}
for $m\in\futon$ (the author does not know whether or not a Hamiltonian 
of this type is ``natural'' in quantum optics or quantum filed theories).

This has a $m$-fold degenerate vacuum because if we set
\begin{equation}
  {\cal C}
  \equiv
  \mathrm{Vect}\left\{\ket{0},\ket{1},\cdots,\ket{m-1}\right\}\ ,
\end{equation}
then $H_0{\cal C}=0$.

Now note that $(\ket{0},\ket{1}\cdots,\ket{m-1})\in\kansu{\stm}{\calh}$
in (\ref{eq:stmh}).
We consider a two-parameter isospectral family
\begin{eqnarray}
  \label{eq:hu}
  &&H_{(\lambda,\mu)}
  \equiv
  \kansu{U}{\lambda,\mu}H_0\kansu{U}{\lambda,\mu}^\dagger\ ,
  \\
  &&\kansu{U}{\lambda,\mu}
  \equiv
  e^{\lambda a^\dagger-\bar\lambda a}e^{\mu K_+-\bar\mu K_-}\ ,
   \label{eq:huni}
\end{eqnarray}
where $(\lambda,\mu)\in\fukuso^2$.
Since (\ref{eq:hu}) is isospectral we have no level-crossing of
eigenvalues for the parameters (adiabatic!).
In the following we focus our attension on the $m$-fold degenerate
vacuum.

$U\equiv U(\lambda,\mu)$ in (\ref{eq:huni}) is unitary, so
\begin{eqnarray}
  &&U\left(\ket{0},\ket{1},\cdots,\ket{m-1}\right)
  =
  \left( U\ket{0},U\ket{1},\cdots,U\ket{m-1}\right)
  \in\kansu{\stm}{\calh}\ ,
  \\
  &&U\left(\sum^{m-1}_{j=0}\ket{j}\bra{j}\right)U^\dagger
  \in\kansu{\grm}{\calh}\ .
\end{eqnarray}
Namely (\ref{eq:hu}) with (\ref{eq:huni}) gives a classifying map
\begin{equation}
  P : \fukuso^2\rightarrow\kansu{\grm}{\calh}\ ,\quad 
  \kansu{P}{\lambda,\mu}
  \equiv
  \kansu{U}{\lambda,\mu}
  \left(\sum^{m-1}_{j=0}\ket{j}\bra{j}\right)
  \kansu{U}{\lambda,\mu}^\dagger
\end{equation}
in our terminology.
From now on our target is the pull-back bundle (\ref{eq:hikimodoshi})
by this map:
\begin{eqnarray}
  \label{eq:ue}
  &&\left(\kansu{U}{m},E,\pi_E,\fukuso^2\right)
  =
  P^*\left(\kansu{U}{m},\kansu{\stm}{\calh},\pi,\kansu{\grm}{\calh}\right)\ ,
  \nonumber\\
  &&E
  =
  \left\{
    \left(
      \left(\lambda,\mu\right),\kansu{U}{\lambda,\mu}
      \left(\ket{0},\ket{1},\cdots,\ket{m-1}\right)
    \right)
    \vert
    \left(\lambda,\mu\right)\in\fukuso^2
  \right\}\ .
\end{eqnarray}
First of all let us calculate a canonical connection form 
(adiabatic connection) (\ref{eq:ateigi}) for (\ref{eq:ue}).
Setting for simplicity
\begin{equation}
  \kansu{V}{\lambda,\mu}
  =
  \kansu{U}{\lambda,\mu}
  \left(\ket{0},\ket{1},\cdots,\ket{m-1}\right)
  \equiv\kansu{U}{\lambda,\mu}V_0\ ,
\end{equation}
the connection form $\cala$ is
\begin{equation}
  \label{eq:setsuzoku}
  \cala
  =
  \kansu{V}{\lambda,\mu}^\dagger\kansu{dV}{\lambda,\mu}
  =
  V_0^\dagger\kansu{U}{\lambda,\mu}^\dagger
  \kansu{dU}{\lambda,\mu}V_0\ ,
\end{equation}
where 
\[
  d
  =
  d\lambda{\partial\over\partial\lambda}
  +d\mu{\partial\over\partial\mu}
  +d\bar\lambda{\partial\over\partial\bar\lambda}
  +d\bar\mu{\partial\over\partial\bar\mu}\ .
\]
To calculate $U^\dagger dU$ we utilize (\ref{eq:ktachi}) and
(\ref{eq:bunkai}).
Making use of
\begin{eqnarray}
  U
  \equiv
  \kansu{U}{\lambda,\mu}
  &=&
  e^{\lambda a^\dagger-\bar\lambda a}
  e^{(\mu(a^\dagger)^2-\bar\mu a^2)/2}
  \nonumber\\
  &=&
  e^{-\vert\lambda\vert^2/2}e^{\lambda a^\dagger}e^{-\bar\lambda a}
  e^{(\mu(a^\dagger)^2-\bar\mu a^2)/2}\ ,\  \mathrm{or}\\
  &=&
  e^{\lambda a^\dagger-\bar\lambda a}
  e^{\zeta(a^\dagger)^2/2}
  e^{\kansu{\log}{1-\vert\zeta\vert^2}{1\over2}(a^\dagger a+{1\over2})}
  e^{-\bar\zeta a^2/2}\ ,
\end{eqnarray}
where $\zeta=\mu\tanh\muzeta/\muzeta$, 
we can calculate $U^{-1}\partial_\lambda U$ and 
$U^{-1}\partial_\mu U$.
Before stating our calculation, we list some useful formulas:
\begin{eqnarray}
  &&\partial_z \left( z{\tanh\vert z\vert\over\vert z\vert}\right)
  =
  {1\over2}
  \left(1-\tanh^2\vert z\vert+{\tanh\vert z\vert\over\vert z\vert}\right)\ ,
  \nonumber\\
  &&\partial_z\kansu{\log}{1-\tanh^2\vert z\vert}
  =
  -{\bar z\tanh\vert z\vert\over\vert z\vert}\ ,
  \nonumber\\
  &&\partial_z
  \left(\bar z{\tanh\vert z\vert\over\vert z\vert}\right)
  =
  {{\bar z}^2\over2\vert z\vert^2}
  \left(1-\tanh^2\vert z\vert-{\tanh\vert z\vert\over\vert z\vert}\right)\ .
  \nonumber
\end{eqnarray}
Let us state our result.

\noindent{\bfseries Lemma 1} We have
\begin{eqnarray}
  \label{eq:lemmnaichinoue}
  U^{-1}\partial_\lambda U
  &=&
  {\bar\lambda\over2}1+\cosh\muzeta a^\dagger
  +{\bar\mu\sinh\muzeta\over\muzeta}a\ ,
  \\
  U^{-1}\partial_\mu U
  &=&
  {1\over4}
  \left(1+{\cosh\muzeta\sinh\muzeta\over\muzeta}\right)
  \left( a^\dagger\right)^2
  \nonumber\\
  &&+{1\over2}
  {\bar\mu\sinh^2\muzeta\over\muzeta^2}
  \left( a^\dagger a+{1\over2}\right)
  \nonumber\\
  &&+{1\over4}
  {\bar\mu^2\over\muzeta^2}
  \left(-1+{\cosh\muzeta\sinh\muzeta\over\muzeta}\right)
  a^2\ .
  \label{eq:lemmnaichinoshita}
\end{eqnarray}
Compare (\ref{eq:lemmnaichinoue}) and (\ref{eq:lemmnaichinoshita})
with those of \cite{ZM1}.
Since the connection $\cala$ is anti-hermitian ($\cala^\dagger=-\cala$), 
it can be written as
\begin{equation}
  \label{eq:calaa}
  \cala
  =
  A_\lambda d\lambda+A_\mu d\mu-A_\lambda^\dagger d\bar\lambda
  -A_\mu^\dagger d\bar\mu\ ,
\end{equation}
so we have, for $\kappa=\lambda, \mu$,
\begin{equation}
  A_\kappa
  =
  V_0^\dagger U^{-1}\partial_\kappa UV_0
  =
  \left(\bra{i}U^{-1}\partial_\kappa U\ket{j}\right)\ ,\
  0\le i, j\le m-1\ ,
\end{equation}
comparing (\ref{eq:calaa}) with (\ref{eq:setsuzoku}).

Now it is easy to find $A_\lambda$ and $A_\mu$ using Lemma 1 and 
(\ref{eq:shoukou}).

\noindent{\bfseries Proposition 2}
We have 
\begin{eqnarray}
  \label{eq:alambda}
  A_\lambda
  &=&
  \small
  \pmatrix{
    {\bar\lambda\over2}&{\bar\mu\sinh\muzeta\over\muzeta}&&&\cr
    \cosh\muzeta&{\bar\lambda\over2}&
    \sqrt{2}{\bar\mu\sinh\muzeta\over\muzeta}&&0&\cr
    &\sqrt{2}\cosh\muzeta&\ddots&\ddots&&\cr
    &&\ddots&\ddots&\ddots&\cr
    &0&&\ddots&{\bar\lambda\over2}&\
    \sqrt{m-1}{\bar\mu\sinh\muzeta\over\muzeta}\cr
    &&&&\sqrt{m-1}\cosh\muzeta&{\bar\lambda\over2}\cr
    }
  \\
  A_\mu
  &=&
  \scriptsize
  \pmatrix{
    {1\over2}\alpha&0&\sqrt{2}\beta&&&&\cr
    0&({1\over2}+1)\alpha&0&\sqrt{6}\beta&&0&\cr
    \sqrt{2}\gamma&0&({1\over2}+2)\alpha&&&&\cr
    &\sqrt{6}\gamma&&&&&\cr
    &&&&\ddots&&\cr
    &&&&&0&\sqrt{(m-2)(m-1)}\beta\cr
    &0&&&0&({1\over2}+m-2)\alpha&0\cr
    &&&&\sqrt{(m-2)(m-1)}\gamma&0&({1\over2}+m-1)\alpha\cr
    }
  \label{eq:amu}
\end{eqnarray}
where $\alpha, \beta$ and $\gamma$ are, respectively,
\begin{equation}
  \gamma
  \equiv
  {1\over4}
  \left(1+{\cosh\muzeta\sinh\muzeta\over\muzeta}\right),
  \nonumber\\
  \alpha
  \equiv
  {1\over2}{\bar\mu\sinh^2\muzeta\over\muzeta^2},
  \nonumber\\
  \beta
  \equiv
  {\bar\mu^2\over4\muzeta^2}
  \left(-1+{\cosh\muzeta\sinh\muzeta\over\muzeta}\right).
  \nonumber
\end{equation}


A comment here is in order.
By the diagonal parts of (\ref{eq:alambda}) and (\ref{eq:amu})
we have the Berry phase stated in \cite{SLB} easily.

Since we have obtained the adiabatic connection form $\cala$, let us calculate 
the curvature forn $\calf$ in (\ref{eq:fteigi}).
A little calculation with (\ref{eq:calaa}) leads to
\begin{eqnarray}
  \label{eq:ften}
  \calf
  &=&
  \left(
    \partial_\lambda A_\mu-\partial_\mu A_\lambda
    +[A_\lambda,A_\mu]
  \right) d\lambda\wedge d\mu
  \nonumber\\
  &&
  -\left(
    \partial_\lambda A_\lambda^\dagger+\partial_{\bar\lambda}A_\lambda
    +[A_\lambda,A_\lambda^\dagger]
  \right) d\lambda\wedge d\bar\lambda
  \nonumber\\
  &&
   -\left(
    \partial_\lambda A_\mu^\dagger+\partial_{\bar\mu}A_\lambda
    +[A_\lambda,A_\mu^\dagger]
  \right) d\lambda\wedge d\bar\mu
  \nonumber\\
  &&-\left(
    \partial_\mu A_\lambda^\dagger+\partial_{\bar\lambda}A_\mu
    +[A_\mu,A_\lambda^\dagger]
  \right) d\mu\wedge d\bar\lambda
  \nonumber\\
  &&-\left(
    \partial_\mu A_\mu^\dagger+\partial_{\bar\mu}A_\mu
    +[A_\mu,A_\mu^\dagger]
  \right) d\mu\wedge d\bar\mu
  \nonumber\\
  &&-\left(
    \partial_{\bar\lambda}A_\mu^\dagger-\partial_{\bar\mu}A_\lambda^\dagger
    -[A_\lambda^\dagger,A_\mu^\dagger]
  \right) d\bar\lambda\wedge d\bar\mu
  \nonumber\ ,
\end{eqnarray}
To calculate each term in (\ref{eq:ften}) let us introduce some notations.
We set $E, F, K, L\in M(m;\fukuso)$ such as
\begin{eqnarray}
  \label{eq:efkl}
  E
  &=&
  \pmatrix{
    0&1&&&&\cr
    &0&\sqrt{2}&&&\cr
    &&0&\ddots&&\cr
    &&&\ddots&\ddots&\cr
    &&&&0&\sqrt{m-1}\cr
    &&&&&0\cr
    }\ ,\
  F=E^\dagger\ ,
  \nonumber\\
  K
  &=&
  \pmatrix{
    0&&&&\cr
    &\ddots&&&\cr
    &&0&&\cr
    &&&0&\cr
    &&&&1\cr
    }\ ,\
  L=
  \pmatrix{
    0&&&&\cr
    &\ddots&&&\cr
    &&0&&\cr
    &&&1&\cr
    &&&&1\cr
    }\ .
\end{eqnarray}
Note that
\begin{equation}
  EK
  =
  \pmatrix{
    0&0&&&&\cr
    &0&0&&&\cr
    &&\ddots&\ddots&&\cr
    &&&&0&\sqrt{m-1}\cr
    &&&&&0\cr
    }\ ,\
  KF=(EK)^\dagger\ .
\end{equation}
Now we state our calculation.

\noindent{\bfseries Proposition 3}
\begin{eqnarray}
  \calf
  &=&
  \Bigg\{
    m{{\bar\mu}^2\cosh\muzeta\over4\muzeta^2}
    \left(
      -1+{\cosh\muzeta\sinh\muzeta\over\muzeta}
    \right) EK
    \nonumber\\
    &&
    \ \ \ \ \ -m{\bar\mu\sinh\muzeta\over4\muzeta}
    \left(
      1+{\cosh\muzeta\sinh\muzeta\over\muzeta}
    \right) KF
  \Bigg\}
  d\lambda\wedge d\mu
  \nonumber\\
  &&
  -mK d\lambda\wedge d\bar\lambda
  \nonumber\\
  &&
  -\Bigg\{
  m{\cosh\muzeta\over4}
  \left(1+{\cosh\muzeta\sinh\muzeta\over\muzeta}\right) EK
  \nonumber\\
  &&
 \ \ \ \ \ \ -m{\mu\sinh\muzeta\over4\muzeta}
  \left(-1+{\cosh\muzeta\sinh\muzeta\over\muzeta}\right) KF
  \Bigg\} d\lambda\wedge d\bar\mu
  \nonumber\\
  &&
  -\Bigg\{
  -m{\bar\mu\sinh\muzeta\over4\muzeta}
  \left(-1+{\cosh\muzeta\sinh\muzeta\over\muzeta}\right) EK
  \nonumber\\
  &&
  \ \ \ \ \ \ +m{\cosh\muzeta\over4}
  \left(1+{\cosh\muzeta\sinh\muzeta\over\muzeta}\right) KF
  \Bigg\} d\mu\wedge d\bar\lambda
  \nonumber\\
  &&
  -\Bigg\{
  {m\over2}{\cosh\muzeta\sinh\muzeta\over\muzeta}K
  +{m(m-1)\over4}{\cosh\muzeta\sinh\muzeta\over\muzeta}L
  \Bigg\}
  d\mu\wedge d\bar\mu
  \nonumber\\
  &&
  -\Bigg\{
  -m{\mu\sinh\muzeta\over4\muzeta}
  \left(1+{\cosh\muzeta\sinh\muzeta\over\muzeta}\right) EK
  \nonumber\\
  &&
  \ \ \ \ \ \ +m{\mu^2\cosh\muzeta\over4\muzeta^2}
  \left(-1+{\cosh\muzeta\sinh\muzeta\over\muzeta}\right) KF
  \Bigg\}
  d\bar\lambda\wedge d\bar\mu\ .
\end{eqnarray}
This is our main result.
We, in particular, consider the case of $m=2$.
Since
\begin{eqnarray}
  &&
  E=\pmatrix{&1\cr0&\cr}\ ,\
  F=\pmatrix{&0\cr1&\cr}\ ,\
  K=\pmatrix{0&\cr&1\cr}\ ,\
  L=\pmatrix{1&\cr&1\cr}\ ,\
  \nonumber\\
  &&EK=E\ ,\ KF=F\ ,
\end{eqnarray}
it is easy to see that the target of $\calf$ covers all of Lie 
algebra $u(2)$.
This means that the connection $\cala$ is irreducible $\cdots$
the holonomy group of $\cala$ is just $U(2)$.
See \cite{ZM1}, \cite{ZM2} and \cite{MN}.
However for $m\ge3$ the target of $\calf$ does not cover
all of $u(m)$, so $\cala$ is not irreducible.

\noindent{\bfseries Corollary 4}
When $m=2$, $\cala$ is irreducible (\cite{ZM1}),
while $\cala$ is not irreducible for $m\ge3$.

Now since we have obtained the connection form $\calf$,
let us moreover calculate $\calf^2$ ($\calf^k=0$ for $k\ge3$
becomes $\mathrm{dim}_\fukuso\fukuso^2=2$).
A little calculation leads to

\noindent{\bfseries Corollary 5}
\begin{equation}
  \calf^2
  =
  \Bigg\{
  {m^2(m-1)\over4}{\cosh\muzeta\sinh\muzeta\over\muzeta}L
  -{m^2(m+1)\over2}{\cosh\muzeta\sinh\muzeta\over\muzeta}K
  \Bigg\}
  d\lambda\wedge d\mu\wedge d\bar\lambda\wedge d\bar\mu\ .
\end{equation}
We have obtained only abelian parts of Lie algebra $u(m)$.
We are now in a stage to calculate several geometric quantities 
such as Chern class, Chern character and Chern-Simons class
(see \cite{MN}) making use of $\cala$ (Proposition 2),
$\calf$ (Proposition 3) and $\calf^2$ (Corollary 5).
However we leave these calculations to the (young!) readers because
they are good excercises to learn the geometric method in 
mathematical physics.

\vspace{5mm}

We would like to close this paper by proposing a future subject.
From Corollary 4 the connection form $\cala$ is not irreducible
for $m\ge3$.
This is insufficient for ``geometric'' quantum computation (\cite{ZM1}),
so that we must make a further generalization of our model.
For example, for the Hamiltonian (\ref{eq:ham}) we would like to consider
a $m$-parameter isospectral family:
\begin{eqnarray}
  &&
  H_{(\lambda_1,\cdots,\lambda_m)}
  \equiv
  \kansu{U}{\lambda_1,\cdots,\lambda_m}H_0
  \kansu{U}{\lambda_1,\cdots,\lambda_m}^\dagger\ ,
  \\
  &&
  \kansu{U}{\lambda_1,\cdots,\lambda_m}
  \equiv
  \calp\prod^m_{j=1}\exp
  \left\{
    (\lambda_j(a^\dagger)^j-\bar\lambda_ja^j)/j
  \right\}\ ,
  \label{eq:ulm}
\end{eqnarray}
where $(\lambda_1,\cdots,\lambda_m)\in\fukuso^m$ and $\calp$ means
path-ordering.
This model may be good at first sight.
However we meet a difficulty immediately.
Since a disentangling formula such as (\ref{eq:ktachi})
or (\ref{eq:kdaisu}) is not known as far as we know,
we can not calculate the connection form $\cala$ from (\ref{eq:ulm}).
As for disentangling formulas see \cite{FS} or \cite{NT}.
It is an important subject to overcome this difficulty.

\vspace{5mm}

\noindent{\em Acknowledgement.}\\
The author wishes to thank Dr. K. Funahashi for his helpful comments and  
suggestions.


%
\end{document}